\documentclass[10pt]{article}
\usepackage[centertags]{amsmath}
\usepackage{amsfonts}
\usepackage{amssymb}
\usepackage{amsthm}
\usepackage{newlfont}
\usepackage{epsfig}
\usepackage{amscd}

\theoremstyle{plain}
\newtheorem{Th}{Theorem}[section]
\newtheorem{Cor}[Th]{Corollary}

\newtheorem{Prop}[Th]{Proposition}
\theoremstyle{definition}
\newtheorem{Def}{Definition}[section]

\theoremstyle{remark}
\newtheorem*{Rem}{Remark}%[section]
\numberwithin{equation}{section}
\newcommand{\PP}{{\mathbb P}}
\newcommand{\RR}{{\mathbb R}}
\newcommand{\EE}{{\mathbb E}}
\newcommand{\ZZ}{{\mathbb Z}}

\newcommand{\cL}{{\mathcal L}}

\newcommand{\bx}{{\boldsymbol x}}

\newcommand{\by}{{\boldsymbol y}}
\newcommand{\bq}{{\boldsymbol q}}
\newcommand{\bp}{{\boldsymbol p}}
\newcommand{\bu}{{\boldsymbol u}}

\begin{document}

\title{Geometric discretization of the Koenigs nets}

\author{Adam Doliwa \\ 
{\it Instytut Fizyki Teoretycznej, Uniwersytet Warszawski}\\
{\it ul. Ho\.{z}a 69, 00-681 Warszawa, Poland}\\ 
e-mail: {\tt doliwa@fuw.edu.pl}}
\date{}
\maketitle

\begin{abstract}

\noindent We introduce the Koenigs lattice, which is a new integrable 
reduction of the quadrilateral lattice (discrete conjugate net)
and provides natural integrable discrete analogue of the
Koenigs net. We construct the Darboux-type transformation 
of the Koenigs lattice and we show permutability of superpositions of such 
transformations, thus proving integrability of 
the Koenigs lattice. 
We also investigate the geometry of the discrete Koenigs transformation. In
particular we characterize the Koenigs transformation in
terms of an involution determined by a congruence conjugate to the lattice.
\\

{\it 2001 PACS:} 02.30.Ik, 02.40.Dr, 05.45.Yv, 04.60.Nc, 02.40.Hw,

\end{abstract}

\section{Introduction}

In the XIX-th century one of the most favorite subjects of the differential 
geometry 
\cite{Bianchi,DarbouxIV} was investigation of special classes of 
surfaces (or, more appropriate, coordinate systems on surfaces and 
submanifolds) which allow for transformations
exhibiting the so called {\em permutability property}.
Such transformations called, depending on the
context, the Darboux, Bianchi, B\"acklund, Laplace, Moutard, Koenigs,
Combescure, L\'evy, Ribaucour or the fundamental transformation of Jonas,
can be also described in terms of certain families of lines called line
congruences~\cite{Eisenhart-TS,Finikov}.
It turns out that most of
the "interesting" submanifolds is provided by
reductions of conjugate nets, and
the transformations between such submanifolds are the corresponding 
reductions of the fundamental (or Jonas) transformation of the
conjugate nets. 

From the other side such submanifolds are described by solutions of certain
nonlinear partial differential equations, which turn out to be 
extensively studied in the modern
theory of integrable systems; here also the existence of transformations
(called in this context the Darboux transformations) appears to be
essential.
For example, the conjugate nets, their iso-conjugate deformations
and transformations are described~\cite{DMMMS} by the so called
multicomponent Kadomtsev--Petviashvilii hierarchy, which is
considered often as the basic system of equations of the soliton theory
\cite{DKJM,KvL}. 

Recently the integrable discrete
(difference) versions of integrable differential equations attracted a lot
of attention
(see, for example, articles in \cite{SIDEI,SIDEII,SIDEIII,BobenkoSeiler}). The
interest in discrete integrable systems is stimulated from various
directions, like numerical methods, theory of special functions, but also
from statistical and quantum physics \cite{JM,KBI}. The 
discrete integrable systems are considered more fundamental then the
corresponding differential systems. 
Discrete equations include the
continuous theory as the result of a limiting procedure, moreover different
limits may give from one discrete equation various differential ones. 
Furthermore,
discrete equations reveal some symmetries lost in the continuous limit.

During last few years the connection between geometry and integrability
has been observed also at a discrete level. The present paper is the next
one in the series of attempts to construct the integrable discrete
geometry --- the theory of lattice submanifolds described
by integrable difference equations. 

The natural discrete analogs of certain coordinate systems
on surfaces were studied by Sauer~\cite{Sauer}. In particular, he introduced
the discrete conjugate nets in $\EE^3$ as lattices with planar elementary
quadrilaterals. The importance of the discrete conjugate nets 
(the quadrilateral lattices) in the soliton theory was
recognized recently in~\cite{DCN,MQL}. 
The Darboux-type transformations of the quadrilateral lattices have been
found in~\cite{MDS}, and
the geometry of these transformations was investigated in detail 
in~\cite{TQL}. In the literature \cite{DMS,KoSchief2,q-red,DS-sym,W-cong}
there are known various integrable reductions of the quadrilateral lattices.
We introduce here the discrete analogue of the Koenigs reduction of
conjugate nets. Let us state briefly the main definitions, ideas
and results of this paper.

\begin{figure}
\begin{center}
\label{fig:conj-net-K}
\epsffile{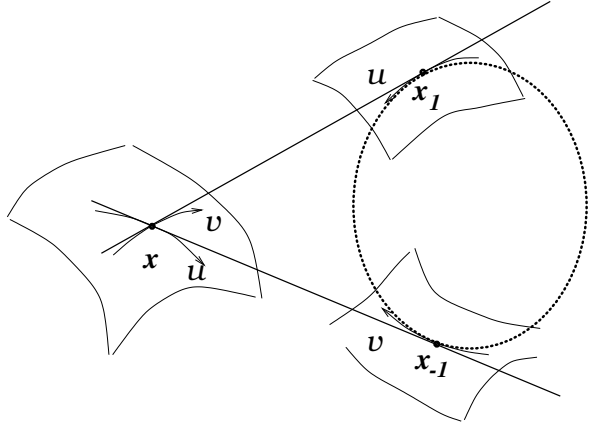}
\caption{The Koenigs net}
\end{center}
\end{figure}

Consider generic two-dimensional conjugate net \cite{Lane} in 
$M$-dimensional projective space $\PP^M$. The homogeneous coordinates
$\bx(u,v)\in\RR^{M+1}_*$ of the net satisfy the Laplace equation
\begin{equation} \label{eq:Laplace-uv}
\bx_{,uv}=a \bx_{,u} + b \bx_{,v} + c \bx,
\end{equation}
where comma denotes differentiation (e.g., 
$\bx_{,u}=\frac{\partial \bx}{\partial u}$), and $a$, $b$, $c$ are functions 
of the conjugate
parameters $(u,v)$ of the net. Its Laplace transforms 
\begin{equation}
\bx_1 = \bx_{,v} - a\bx, \qquad \bx_{-1}=\bx_{,u} -b\bx,
\end{equation} 
are another conjugate nets such that the $v$-tangents of $\bx$ coincide with the
corresponding $u$-tangent lines of $\bx_{1}$ and that the $u$-tangents of 
$\bx$ coincide with the corresponding $v$-tangent lines of $\bx_{-1}$ (see
figure \ref{fig:conj-net-K}). 
In the tangent plane at a point $\bx$ there is a linear system (pencil)
of conics 
tangent to the $u$-coordinate line at the point $\bx_{1}$ and tangent to the
$v$-curve at the point $\bx_{-1}$. When there is one conic of this pencil 
with the second order contact with the $u$-curve of $\bx_{1}$ and with
the second order contact with the $v$-curve of $\bx_{-1}$ then the net is 
called {\em the net of Koenigs}. It turns
out that the Laplace equation \eqref{eq:Laplace-uv} of the Koenigs net
can be gauged into the form
\begin{equation} \label{eq:Koenigs-uv}
\bx_{,uv}=f \bx.
\end{equation}

The integrable discrete analogue of two-dimensional
conjugate net is a $\ZZ^2$-lattice made of
planar quadrilaterals \cite{Sauer,DCN}.
One can construct for such lattices \cite{DCN} the analogue of the Laplace 
transforms $\bx_{1}$ and $\bx_{-1}$.
In the plane of the elementary
quadrilateral at a point $\bx(n_1,n_2)$ there is a pencil of conics 
passing through the points $\bx_{1}(n_1,n_2)$ and  
$\bx_{1}(n_1+1,n_2)$ of the Laplace
transform $\bx_{1}$ and passing through the points $\bx_{-1}(n_1,n_2)$ 
and $\bx_{-1}(n_1,n_2+1)$ of the Laplace transform $\bx_{-1}$; we have
replaced the tangency to a curve by its natural discrete analogue
of passing through two neighbouring points of the
discrete parametric curve. When there is one conic 
of this pencil passing through $\bx_{1}(n_1+2,n_2)$ and passing through 
$\bx_{-1}(n_1,n_2+2)$ then we call such a lattice {\em the Koenigs lattice}. 

The reduction of the fundamental transformation of conjugate nets to the
class of the Koenigs nets is called {\em the transformation of Koenigs}.
Such transformation is determined only by the half of the data of the 
fundamental
transformation: given congruence conjugate to the Koenigs net then the
second Koenigs net $\bx^\prime$ is the harmonic conjugate of $\bx$ with
respect to the focal nets of
the congruence. The discrete analogue of this construction is more
subtle. One can show that $\bx^\prime$ is the image of $\bx$ in an
involution on the corresponding line of the congruence. This
involution is uniquely defined by the focal nets of the congruence;
in the continuous case the focal points are the double
points of the involution \cite{Eisenhart-TS,Lane}

We have sketched the main ideas and results of the paper. The detailed
presentation will be as follows. In Section \ref{sec:K-lattice} we define
and study in detail the Koenigs lattice. In particular we discuss the 
integrability of the Koenigs
lattice from the point of view of the Pascal theorem. 
In Section \ref{sec:K-transformation} we present
the discrete analogue of the Koenigs transformation and then in Section 
\ref{sec:K-geometry} we investigate geometric properties
of the Koenigs transformation. Finally, in Section 
\ref{sec:K-superposition} we investigate superpositions of the Koenigs
transformation and prove their permutability, thus showing
integrability of the Koenigs lattice. 

\section{The Koenigs lattice} \label{sec:K-lattice}
Consider a two-dimensional quadrilateral lattice in $M$-dimensional 
projective
space $\PP^M$, whose points labelled by two-dimensional integer
lattice $\ZZ^2$, satisfy the property of planarity of elementary
quadrilaterals \cite{Sauer,DCN}. In terms of the homogeneous coordinates 
such a lattice is described by solution of the discrete Laplace equation 
\begin{equation} \label{eq:Laplace-d}
\bx_{(12)}=A_{(1)}\bx_{(1)} + B_{(2)}\bx_{(2)} + C\bx,
\end{equation}
where $\bx:\ZZ^2\to\RR^{M+1}_*$ and subscripts in brackets mean shifts along
the $\ZZ^2$ lattice, i.e., 
$\bx_{(\pm 1)}(n_1,n_2)= \bx(n_1\pm1,n_2)$,  
$\bx_{(\pm 2)}(n_1,n_2)= \bx(n_1,n_2\pm 1)$, and
$\bx_{(\pm 1 \pm 2)}(n_1,n_2)= \bx(n_1\pm 1,n_2\pm 1)$. Here also 
$A$, $B$ and $C$ are functions on $\ZZ^2$ which characterize the
lattice completely up to initial curves $\bx(n_1,0)$ and $\bx(0,n_2)$. 
Notice that multiplication of $\bx$ by a non-zero function $\rho$
implies the corresponding change of $A$, $B$ and $C$ but does not change the
lattice itself.

\begin{figure}
\begin{center}
\label{fig:Lapl-transf-d-K}
\epsffile{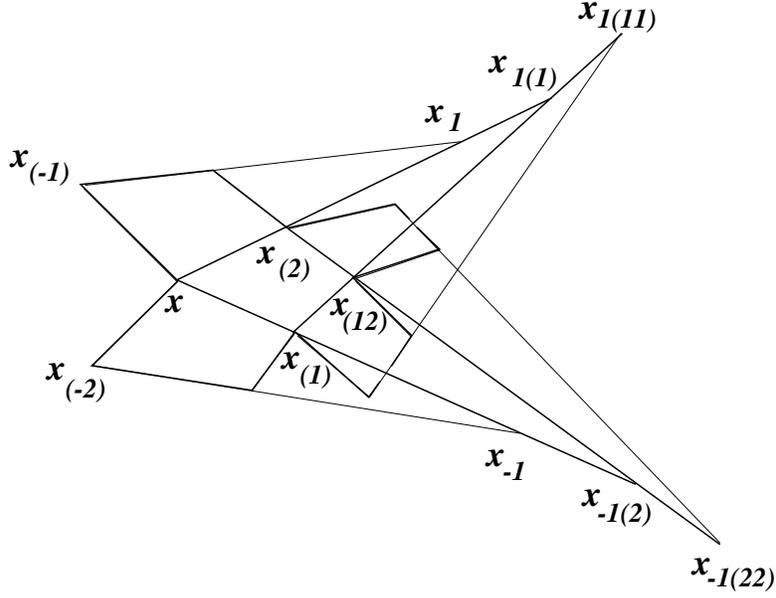}
\caption{The quadrilateral lattice and its Laplace transforms}
\end{center}
\end{figure}

As it was shown in \cite{DCN} because of the planarity of the elementary
quadrilaterals of the lattice one can define its Laplace transforms $\bx_1$
and $\bx_{-1}$ (see figure \ref{fig:Lapl-transf-d-K})

\begin{equation} 
\bx_1=\bx_{(2)}-A\bx, \qquad \bx_{-1}=\bx_{(1)}-B\bx.
\end{equation}
\begin{Rem}
Notice that our notation differs from that of \cite{DCN} by opposite
ordering of the
transformations and by shifts in parameters. Moreover in \cite{DCN} we used the
affine gauge which made some formulas more complicated.
\end{Rem}

Using the discrete Laplace equation \eqref{eq:Laplace-d} one can show that 
$\bx_{1(1)}$ is collinear with $\bx$,
$\bx_{(2)}$ and $\bx_{1}$
\begin{equation}
\bx_{1(1)}=B_{(2)}\bx_{(2)} + C\bx = B_{(2)}\bx_{1} + (B_{(2)}A +C)\bx;
\end{equation}
similarly $\bx_{-1(2)}$ is collinear with $\bx$,
$\bx_{(1)}$ and $\bx_{-1}$
\begin{equation}
\bx_{-1(2)}=A_{(1)}\bx_{(1)} + C\bx = A_{(1)}\bx_{-1} + (A_{(1)}B +C)\bx.
\end{equation}
\begin{Rem}
Two functions $H$ and $K$ defined \cite{DCN} as the cross-ratios
\begin{eqnarray}
H &= cr(\bx_{(1)},\bx;\bx_{-1},\bx_{-1(2)}) = -\frac{A_{(1)}B}{C}, \\
K &= cr(\bx_{(2)},\bx;\bx_{1},\bx_{1(1)}) = -\frac{B_{(2)}A}{C},
\end{eqnarray}
are gauge-invariant and are
called the invariants of the lattice $\bx$. They are natural discrete
analogues of the invariants
\begin{equation}
h=c+ab-a_{,u}, \qquad k=c+ab-b_{,v},
\end{equation}
of conjugate nets.
In the continuous case the Koenigs nets have equal invariants, i.e., $h=k$.
This property does not transfer to the discrete case.
\end{Rem}

It is well known (see for example \cite{Samuel})
that five distinct points in a projective plane, no four of which are 
collinear, uniquely determine a conic. Moreover, a pencil of conics 
(one-dimensional linear subspace of the five dimensional
space of conics) is uniquely determined by four points (the base of the
pencil) no three of which are collinear. 

The four points $\bx_{1}$, $\bx_{1(1)}$, $\bx_{-1}$ and $\bx_{-1(2)}$
belong to the plane $P_{\bx\bx_{(1)}\bx_{(2)}}$ of the elementary
quadrilateral of $\bx$ and define a linear system of conics. Let us choose
the points $\bx$, $\bx_{-1(2)}$, $\bx_{1(1)}$ as vertices of the local
triangle of reference in that plane, i.e., a point
$y_1\bx+y_2\bx_{-1(2)}+y_3\bx_{1(1)}$ has coordinates proportional to
$(y_1,y_2,y_3)$. Then the equation of a general conic of the pencil is of
the form
\begin{equation} \label{ref:pencil-K}
y_1^2 + (A_{(1)}B+C)y_1 y_2 + (B_{(2)}A +C)y_1y_3 + \lambda y_2 y_3 = 0,
\end{equation}
with $\lambda$ being a parameter. 

\begin{Def}
{\em The Koenigs lattice} is a two dimensional quadrilateral lattice such that for
every point $\bx$ of the lattice there exist a conic passing through the six
points $\bx_{1}$, $\bx_{1(1)}$,$\bx_{1(11)}$, 
$\bx_{-1}$, $\bx_{-1(2)}$ and $\bx_{-1(22)}$.
\end{Def}

\begin{Prop}
The Laplace equation of the 
Koenigs lattice can be gauged into the canonical form 
\begin{equation} \label{eq:Koenigs-d}
\bx_{(12)} + \bx =F_{(1)}\bx_{(1)} + F_{(2)}\bx_{(2)}.
\end{equation}
\end{Prop}
\begin{proof}

The points $\bx_{1(11)}$ and $\bx_{-1(22)}$ also belong to the plane
$P_{\bx\bx_{(1)}\bx_{(2)}}=P_{\bx\bx_{-1(2)}\bx_{1(1)}}$ and have the
following decompositions  
\begin{eqnarray}
\bx_{1(11)}=-\left( CB_{(12)} + \frac{C C_{(1)}}{A_{(1)}} \right)\bx +
\left( B_{(12)} + \frac{C_{(1)}}{A_{(1)}} \right)\bx_{-1(2)} +
B_{(12)}\bx_{1(1)}, \nonumber \\
\bx_{-1(22)}= -\left( C A_{(12)} +\frac{C C_{(2)}}{B_{(2)}} \right) \bx +
A_{(12)}\bx_{-1(2)} +
\left( A_{(12)} + \frac{C_{(2)}}{B_{(2)}}\right) \bx_{1(1)}. \nonumber
\end{eqnarray}
In the pencil \eqref{ref:pencil-K} there exist a conic passing through
$\bx_{1(11)}$ and $\bx_{-1(22)}$ if and only if the coefficients of the 
Laplace equation \eqref{eq:Laplace-d}  
of the Koenigs lattice satisfy the
constraint
\begin{equation} \label{eq:constr-ABC}
\frac{A C_{(2)}}{A_{(12)}}=\frac{B C_{(1)}}{B_{(12)}}. 
\end{equation}
This constraint is
the compatibility condition of the linear system for the unknown function 
$\rho$
\begin{eqnarray}
\rho_{(12)}&=&-C\rho, \label{eq:rho-1}\\
\rho_{(1)}A&=&\rho_{(2)}B \label{eq:rho-2}.
\end{eqnarray}
Using solution of this system as the gauge function we obtain 
new representation
\begin{equation}
\tilde\bx=\frac{1}{\rho}\bx ,
\end{equation}
of the Koenigs lattice which satisfies equation \eqref{eq:Koenigs-d} with
\begin{equation} \label{eq:newF-K}
F=\frac{A\rho}{\rho_{(2)}}=\frac{B\rho}{\rho_{(1)}}.
\end{equation} 
\end{proof}
\begin{Rem}
Equation \eqref{eq:Koenigs-d}, in a gauge equivalent form, appeared first 
in \cite{NDS,DNS-I} in connection with the integrable discretization of the 
Bianchi--Ernst system.
\end{Rem}
Usually the integrability of a nonlinear problem is connected with its 
hidden linear structure. It turns out that, with the help of the celebrated
Pascal theorem on six points on a conic, the discrete Koenigs
constraint can be formulated in a linear way. 

\begin{Prop}
The quadrilateral lattice $\bx$ is the Koenigs lattice if and only if the
lines $L_{\bx_{-1}\bx_{1(11)}}$, $L_{\bx_{1}\bx_{-1(22)}}$ and
$L_{\bx \bx_{(12)}}$ intersect in a single point.
\end{Prop}
\begin{proof}

Recall that given six point $1$, $2$, $3$, $4$, $5$, $6$ belong to a conic 
if and only if the points
$i=L_{12}\cap L_{45}$, $j=L_{23}\cap L_{56}$ and $k=L_{34}\cap L_{61}$
are collinear.
We apply the Pascal theorem to the
six points $\bx_{-1}$, $\bx_{-1(2)}$, $\bx_{-1(22)}$, $\bx_{1}$, 
$\bx_{1(1)}$ and
$\bx_{1(11)}$. 
Because point $\bx$ is the intersection of lines $L_{\bx_{-1}\bx_{-1(2)}}$ 
and 
$L_{\bx_{1}\bx_{1(1)}}$ and the point $\bx_{(12)}$ is the intersection of 
lines $L_{\bx_{-1(2)}\bx_{-1(22)}}$ and $L_{\bx_{1(1)}\bx_{1(11)}}$ then 
there
exists a conic passing through the six points if and only if the
statement of the proposition holds. 
\end{proof}

\section{The discrete Koenigs transformation} \label{sec:K-transformation}
The Koenigs transformation is the reduction of the fundamental
transformation to the class of the Koenigs nets and lattices. Let us first
recall relevant definitions and results from the theory of transformations
of quadrilateral lattices \cite{TQL}. Then we present the algebraic
definition of the Koenigs reduction of the fundamental transformation. We
postpone to next section the discussion of the geometric 
interpretation of the Koenigs transformation.

\subsection{The fundamental transformation of quadrilateral lattices}
We recall the basic results from the theory of transformations of
quadrilateral lattices \cite{TQL}. The novelty here is the description of 
the theory in the homogeneous formalism (but the geometric content does
not change). We constraint our presentation to two dimensional lattices 
and congruences only. 

\begin{Def}
The discrete two-dimensional line congruence $L$ is a $\ZZ^2$-family of 
lines in
$\PP^M$ such that any two neighbouring lines intersect. The intersection
$\by_i=L\cap L_{(-i)}$, $i=1,2$, is called the $i$-th focal lattice of the
congruence.
\end{Def}
\begin{Cor}
The focal lattices of discrete two-dimensional line congruences are
quadrilateral lattices.
\end{Cor}
\begin{Def} A two-dimensional quadrilateral lattice $\bx$ and a 
two-dimensional congruence $L$ are called conjugate 
when points of the lattice belong to the corresponding lines of the
congruence, i.e.,  $\bx(n_1,n_2)\in L(n_1,n_2)$ for all 
$(n_1,n_2)\in\ZZ^2$.
\end{Def}
\begin{Def}
The quadrilateral lattice $\bx^\prime$ is a fundamental transform of $\bx$
if there exists a congruence (called the congruence of
the transformation) conjugate to both lattices.
\end{Def}
\begin{Th} \label{th:fund-2-app}
Two quadrilateral lattices $\bx$ and $\bx^\prime$ are fundamental transforms
of each other if and only if there exist solutions $\phi$ and $\phi^\prime$
of the Laplace equations of the lattices and there exist functions $k$ and
$\ell$ such that the system
\begin{eqnarray} \label{eq:fund-x-x'}
\Delta_1 \left( \frac{\bx^\prime}{\phi^\prime}\right) = 
k_{(1)}\Delta_1\left( \frac{\bx}{\phi}\right),\\
\Delta_2 \left( \frac{\bx^\prime}{\phi^\prime}\right) = 
\ell_{(2)}\Delta_2\left( \frac{\bx}{\phi}\right),
\end{eqnarray} 
is satisfied. 
\end{Th}
\begin{Cor}
The system \eqref{eq:fund-x-x'} is compatible if and only if
there exist a solution
$\theta$ of the equation
\begin{equation} \label{eq:Laplace-adj-d}
C\theta_{(12)}= -B\theta_{(1)} -A\theta_{(2)} + \theta,
\end{equation}
called the adjoint of \eqref{eq:Laplace-d},
and the functions $k$ and $\ell$ are solution of the following
system
\begin{eqnarray}
k-\ell&=&\phi\theta, \nonumber\\
\Delta_1\ell &=& -(\phi_{(1)}-B\phi)\theta_{(1)}, \label{eq:kl}\\
\Delta_2k &=& (\phi_{(2)}-A\phi)\theta_{(2)} \nonumber.
\end{eqnarray}
\end{Cor}
\begin{Cor}
The fundamental transformation of the given lattice
$\bx$ can be constructed when we are given a solution of its Laplace
equation and a solution of its adjoint (both are given up to two functions
of single variables). The next step is to find the functions
$k$ and
$\ell$ (given up to a constant)
by solving the system \eqref{eq:kl}. Finally, the transformed 
lattice in the gauge
\begin{equation}
\hat\bx=\frac{\bx^\prime}{\phi^\prime},
\end{equation} 
is obtained (up to a constant vector) from the system \eqref{eq:fund-x-x'}.
The coefficients of the Laplace equation of the lattice $\hat\bx$
read
\begin{equation}
\hat{A}=A\frac{k_{(2)}\phi}{k \phi_{(2)}}, \qquad 
\hat{B}=B\frac{\ell_{(1)}\phi}{\ell \phi_{(1)}}, \qquad \hat{C} = 1 -
\hat{A}_{(1)} - \hat{B}_{(2)}.
\end{equation}
\end{Cor}
\begin{Rem}
The corresponding tangent lines of $\bx$ and $\bx^\prime$ intersect
in points of the quadrilateral lattices
\begin{equation}
\cL_1(\bx) = \Delta_1\left( \frac{\bx}{\phi}\right) , \qquad
\cL_2(\bx) = \Delta_2\left( \frac{\bx}{\phi}\right),
\end{equation} 
called the L\'evy transforms \cite{TQL} of $\bx$.
\end{Rem}
\begin{Cor} \label{cor:focal-y}
The focal lattices of the congruence of the transformation given by
\begin{equation} \label{eq:focal-lat}
\by_1 = k\frac{\bx}{\phi} - \frac{\bx^\prime}{\phi^\prime}, \qquad
\by_2 = \ell\frac{\bx}{\phi} -\frac{\bx^\prime}{\phi^\prime},
\end{equation} 
satisfy equations 
\begin{eqnarray}
\by_1-\by_2&=&\theta\bx, \nonumber\\
\Delta_1\by_2 &=& -(\bx_{(1)}-B\bx)\theta_{(1)}, \label{eq:y1y2}\\
\Delta_2\by_1 &=& (\bx_{(2)}-A\bx)\theta_{(2)} \nonumber,
\end{eqnarray}
i.e., they
can be found using the solution $\theta$ of the adjoint
equation  \eqref{eq:Laplace-adj-d} only.  
\end{Cor}
\begin{Rem}
Equations \eqref{eq:y1y2} can be used to find
congruences conjugate to the
lattice $\bx$.
Notice that the role of the new solution $\phi$ of the Laplace equation
\eqref{eq:Laplace-d} of the lattice $\bx$ in equations \eqref{eq:kl} 
is taken in \eqref{eq:y1y2} by $\bx$ itself. 
\end{Rem}
\begin{Rem}
The lattices $\by_1=\cL^*_1(\bx)$ and $\by_2=\cL^*_2(\bx)$ are also called
the adjoint L\'evy transforms \cite{TQL} of $\bx$.
\end{Rem}

\subsection{The algebraic formulation of the discrete
Koenigs transformation}
\begin{Prop} \label{th:transf-dK}
Given the Koenigs lattice $\bx$ satisfying equation \eqref{eq:Koenigs-d} and
given a scalar solution $\theta$ of its adjoint equation (the Moutard 
equation) 
\begin{equation} \label{eq:Moutard-d}
\theta_{(12)}+\theta = F(\theta_{(1)}+\theta_{(2)}),
\end{equation}
then the solution $\bx^\prime$ of the linear system
\begin{eqnarray}
\Delta_1\left( \frac{\bx^\prime}{\phi^\prime} \right) & = &\;
 (\theta \theta_{(2)})_{(1)}\Delta_1\left( \frac{\bx}{\phi} \right), \\
\Delta_2\left( \frac{\bx^\prime}{\phi^\prime} \right) & = &
- (\theta \theta_{(1)})_{(2)}\Delta_2\left( \frac{\bx}{\phi} \right),
\end{eqnarray}
with 
\begin{equation}
\phi = \theta_{(1)} + \theta_{(2)}, \qquad \phi^\prime = 
\frac{1}{\theta_{(1)}} + \frac{1}{\theta_{(2)}},
\end{equation}
is a new Koenigs lattice satisfying equation \eqref{eq:Koenigs-d} with
\begin{equation} \label{eq:F'}
F^\prime=F\frac{\theta_{(1)}\theta_{(2)}}{\theta\theta_{(12)}}.
\end{equation}
\end{Prop}
\begin{proof}
First one should observe \cite{NDS} that if $\theta$ satisfies the 
Moutard equation \eqref{eq:Moutard-d}
then $\phi=\theta_{(1)}+\theta_{(2)}$ is a solution of 
the Koenigs lattice equation \eqref{eq:Koenigs-d}. Then the corresponding 
solutions of the system
\eqref{eq:kl} are 
\begin{equation}
k=\theta_{(2)}\theta, \qquad \ell=-\theta_{(1)}\theta,
\end{equation}
and the coefficients of the Laplace equation of the new quadrilateral
lattice $\hat\bx$ read
\begin{eqnarray}
\hat{A}&=&F\frac{\theta_{(1)}+\theta_{(2)}}{\theta\theta_{(2)}}
\left(\frac{\theta\theta_{(2)}}{\theta_{(1)}+\theta_{(2)}} \right)_{(2)}, 
\nonumber \\
\hat{B}&=&F\frac{\theta_{(1)}+\theta_{(2)}}{\theta\theta_{(1)}}
\left(\frac{\theta\theta_{(1)}}{\theta_{(1)}+\theta_{(2)}} \right)_{(1)}, 
\\ \hat{C} & = &1 -
\hat{A}_{(1)} - \hat{B}_{(2)}. \nonumber
\end{eqnarray}
The function 
\begin{equation}
\rho=\frac{\theta_{(1)}\theta_{(2)}}{\theta_{(1)}+\theta_{(2)}}
=\frac{1}{\phi^\prime},
\end{equation}
is a solution of the system \eqref{eq:rho-1}-\eqref{eq:rho-2}. This 
implies that
\begin{equation}
\bx^\prime = \phi^\prime\hat\bx,
\end{equation}  
satisfies the Koenigs lattice equation \eqref{eq:Koenigs-d}, and the
corresponding potential, according to equation \eqref{eq:newF-K}, agrees
with that given by \eqref{eq:F'}.
\end{proof}
\begin{Cor}
The function $\phi^\prime$ satisfies the Laplace equation of the lattice
$\bx^\prime$. 
\end{Cor}
\begin{Rem}
The important observation that the adjoint of the
Koenigs lattice equation is the Moutard equation is due to Nieszporski
\cite{Nieszporski-priv-A01}.
\end{Rem}

\section{The geometric meaning of the Koenigs transformation}
\label{sec:K-geometry}

To present the geometric meaning of the discrete Koenigs transformation,
introduced in the previous section, we first recall standard results on 
involutions on a projective line $L$ (see \cite{Samuel}).  

\begin{Th} \label{th:inv-harm-con}
If a projective transformation $h:L\to L$ has two distinct fixed points 
$\bp$
and $\bq$ then $h$ is an involution if and only if for any point $\bu\in L$ its
image $h(\bu)$ is the harmonic conjugate of $\bu$ with respect to $\bp$
and $\bq$. 
\end{Th}
\begin{Th} \label{th:inv-dp}
A projective involution is uniquely determined giving two pairs of 
homologous points.
\end{Th}

\subsection{The discrete Koenigs transformation as geometric reduction of
the fundamental transformation} 
\begin{Prop} \label{prop:invol}
Given Koenigs lattice $\bx$ and its transform $\bx^\prime$, denote by
$\by_1$ and $\by_2$ the focal lattices of the congruence of the
transformation. The Koenigs transform $\bx^\prime$ is the
image of $\bx$ in the unique involution mapping $\by_1$ into $\by_{1(1)}$ 
and $\by_2$ into $\by_{2(2)}$. 
\end{Prop}
\begin{proof}
In the gauge of Proposition \ref{th:transf-dK} and due to Corollary
\ref{cor:focal-y} we have
\begin{equation} \label{eq:x-xK-focal}
\bx=\frac{1}{\theta}\left(\by_1-\by_2\right), 
\qquad \bx^\prime=-\frac{1}{\theta_{(1)}\theta_{(2)}}
\left(\theta_{(1)}\by_1+\theta_{(2)}\by_2\right).
\end{equation}
Equations \eqref{eq:y1y2} imply that
\begin{eqnarray}
\by_{1(1)} & = & \frac{1}{\theta} \left( F\theta_1 \by_1 - (F\theta_1
-\theta)\by_2\right),\\
\by_{2(2)} & = & \frac{1}{\theta} \left( F\theta_2 \by_2 - (F\theta_2
-\theta)\by_1\right).
\end{eqnarray}
The unique involution mapping $\by_1$ into $\by_{1(1)}$ 
and $\by_2$ into $\by_{2(2)}$ is the projection of the linear map, which is
convenient to choose in the form  
\begin{eqnarray}
\by_1&\mapsto& -F \by_1 + \left(F -\frac{\theta}{\theta_1}\right)\by_2 ,\\
\by_2&\mapsto& F \by_2 - \left(F -\frac{\theta}{\theta_2}\right)\by_1 .
\end{eqnarray}
As one can check directly the image of $\bx$ in this 
mapping is $\bx^\prime$.
\end{proof}
\begin{Cor}
In the continuous limit the points of the focal lattices become the double
points of the involution. This fact and Theorem \ref{th:inv-harm-con} imply
that $\bx^\prime$ becomes, in the limit, the harmonic conjugate of $\bx$
with respect to the pair $\by_1$ and $\by_2$.
\end{Cor}
It turns out that the property of the Koenigs transformation described in
Proposition \ref{prop:invol} holds exclusively for the Koenigs lattice 
and selects it from general quadrilateral lattices.

\begin{Prop}
Given two quadrilateral lattices $\bx$ and $\bx^\prime$ related by the
fundamental transformation such that the focal 
lattices $\by_1$, $\by_2$ of the congruence of the transformation do not 
degenerate to a single point. If $\bx^\prime$ is the
image of $\bx$ in the unique involution mapping $\by_1$ into $\by_{1(1)}$ 
and $\by_2$ into $\by_{2(2)}$ then $\bx$ and $\bx^\prime$ are Koenigs
lattices related by the Koenigs transformation.
\end{Prop}
\begin{proof}
Notice first that the excluded situation of the degenerated focal lattices
corresponds to the trivial solution $\theta=0$ of the adjoint equation
\eqref{eq:Laplace-adj-d} of the lattice $\bx$.

Formulas \eqref{eq:focal-lat}  and the assumption of the 
proposition imply the following constraint on the functions $k$ and $\ell$
\begin{equation}
k_{(1)}k = \ell_{(2)}\ell.
\end{equation}
This constraint allows to solve the system \eqref{eq:kl} 
\begin{equation} \label{eq:kl-inv}
k=\frac{\phi\theta\theta_{(2)}A}{A\theta_{(2)}+B\theta_{(1)}},\qquad
\ell=-\frac{\phi\theta\theta_{(1)}B}{A\theta_{(2)}+B\theta_{(1)}},
\end{equation} 
and gives the following relation between the coefficients $A$ and $B$ of the
Laplace equation \eqref{eq:Laplace-d}
\begin{equation}
kB\theta_{(1)} + \ell A \theta_{(2)}=0.
\end{equation}
In consequence we have also
\begin{equation} \label{eq:k1l2-inv}
k_{(1)}= 
-\frac{\phi\theta_{(1)}\theta_{(12)}BC}{A\theta_{(2)}+B\theta_{(1)}},
\qquad
\ell_{(2)}= 
\frac{\phi\theta_{(2)}\theta_{(12)}AC}{A\theta_{(2)}+B\theta_{(1)}}.
\end{equation}
The above relations allow to check that the condition \eqref{eq:constr-ABC}
holds for the lattice $\bx$, which implies that $\bx$ is the Koenigs
lattice (notice that due to the symmetry between both lattices
in the definition of the fundamental transformation and in the notion
of the harmonic conjugate the analogous condition
holds for the lattice $\bx^\prime$).

Assuming therefore that the function $\bx$ of the first lattice
is in the canonical gauge $A=B=F$, $C=-1$, one obtains from equations
\eqref{eq:kl-inv}-\eqref{eq:k1l2-inv} that
\begin{equation}
k=\theta\theta_{(2)}N_2(n_2), \qquad \ell=\theta\theta_{(1)}N_1(n_1),
\end{equation}
where $N_i(n_i)$, $i=1,2$, are functions (still to be determined)
of single variables. Then the function 
\begin{equation}
\phi=\theta_{(1)}N_1(n_1) - \theta_{(2)}N_2(n_2),
\end{equation}
obtained using \eqref{eq:kl}, satisfies equation \eqref{eq:Koenigs-d} for
generic $F$ only if $N_1=-N_2=const.$  Without loss of generality
this constant can be put equal to $1$.  
The rest of the proof is the same like the proof of Proposition
\ref{th:transf-dK}.
\end{proof}
\begin{Rem}
The above proposition in the continuous case was found by Koenigs
\cite{Koenigs}. 
\end{Rem}
\begin{Rem}
The excluded case $\theta=0$ corresponds to the reduction of the fundamental
transformation to the radial transformation \cite{TQL}.
\end{Rem}

\subsection{Further geometric properties of the discrete Koenigs transformation}

To understand more the relation between the Koenigs lattice, as defined
in terms of conics, and the geometric description of 
the Koenigs transformation in terms of involutions on lines of the
congruence, we will need the following result.
\begin{Th}[Desargues---Sturm]
A pencil of conics of a projective plane determines 
a projective involution on every line that does not intersect the base 
of the pencil.
\end{Th}

\begin{Rem}
The results presented in this section are generalization to the discrete 
level of the results of Tzitz\'eica \cite{Tzitzeica-K} and Eisenhart
\cite{Eisenhart-K}.
\end{Rem}

It turns out that the Koenigs transformation defines certain
family of quadrics. This family contains the pencils of conics of the both
Koenigs lattices $\bx$ and $\bx^\prime$. 

\begin{Prop}
Given Koenigs lattice $\bx$ and its transform $\bx^\prime$ then the pencils
of conics of both lattices determine the same involution on the intersection
line of the planes of elementary quadrilaterals of $\bx$ and $\bx^\prime$.
\end{Prop}
\begin{proof}
In the gauge such that $\bx$ satisfies equation \eqref{eq:Koenigs-d}
the equation of the pencil of conics \eqref{ref:pencil-K} reads
\begin{equation} \label{ref:pencil-K-2}
y_1^2 + (F_{(1)}F-1)y_1 y_2 + (F_{(2)}F -1)y_1y_3 + \lambda y_2 y_3 = 0.
\end{equation}
When the parameter $\lambda$ equals
\begin{equation} \label{eq:pencil-K-F}
\lambda_0=\frac{F}{F_{(12)}}+1 - F_{(1)}F - F_{(2)}F,
\end{equation}
the conic passes through $\bx_{1(11)}$ and $\bx_{-1(22)}$.

Instead of the basis $\bx$, $\bx_{-1(2)}$ and $\bx_{1(1)}$ of the plane of 
the elementary quadrilateral of $\bx$ let us choose points $\bx$, $\bu_1$
and $\bu_2$,
where
\begin{equation}
\bu_1 = \Delta_1\left(\frac{\bx}{\phi}\right), \qquad 
\bu_2 = \Delta_2\left(\frac{\bx}{\phi}\right),
\end{equation}
represent points of intersection of the tangent lines of $\bx$ and
$\bx^\prime$. Moreover the line $L_{\bu_1\bu_2}$ is the intersection of 
the planes
of elementary quadrilaterals of $\bx$ and
$\bx^\prime$. The transition formulas read
\begin{eqnarray}
\bx_{-1(2)} & = & F_{(1)}\phi_{(1)} \bu_1 + \bx\left(
\frac{F_{(1)}\phi_{(1)}}{\phi}-1 \right),\\
\bx_{1(1)} & = & F_{(2)}\phi_{(2)} \bu_2 + \bx\left(
\frac{F_{(2)}\phi_{(2)}}{\phi}-1 \right). 
\end{eqnarray}
When $(t_1,t_2,t_3)$ are
coordinates of a point
\begin{equation}
\by = t_1\bx + t_2 \bu_1 + t_3 \bu_2,
\end{equation}
then the equation of the pencil \eqref{eq:pencil-K-F} is transformed into
\begin{equation} \label{eq:pencil-K-3}
t_1^2 + a_1b_1t_2^2 + a_2b_2 t_3^2 + t_1t_2(a_1+b_1) + t_1t_3(a_2+b_2) +
t_2t_3 (a_1b_2 + a_2 b_1 + \mu) = 0,
\end{equation}
where 
\begin{equation}
a_i = \frac{1}{F_{(i)} \phi_{(i)}} - \frac{1}{\phi}, \qquad
b_i= \frac{F}{\phi_{(i)}}-\frac{1}{\phi}, \quad i=1,2,
\end{equation}
and
\begin{equation} \label{eq:mu-lambda}
\mu = \frac{\lambda}{F_{(1)} F_{(2)} \phi_{(1)} \phi_{(2)}}.
\end{equation}

The involution on the line $t_1=0$ is equivalent to the problem of finding
the second root of the quadratic equation
\begin{equation}
a_1b_1t_2^2 + a_2b_2 t_3^2 + t_2t_3 (a_1b_2 + a_2 b_1 + \mu) = 0,
\end{equation}
when the first root is given. The double points of the involution 
correspond to
\begin{equation}
\mu = -(a_1b_2 + a_2b_1) \pm 2\sqrt{a_1b_1a_2b_2},
\end{equation}
and their coordinates are
solutions of the equation
\begin{equation}
(\sqrt{a_1b_1}t_2 \pm \sqrt{a_2b_2}t_3 )^2=0.
\end{equation}
Finally, the double points are given by
\begin{equation}
\sqrt{a_1b_1}\bu_2 \pm \sqrt{a_2b_2}\bu_1.
\end{equation}

To find the equation of the second pencil of conics we 
take the primed version of equation \eqref{eq:pencil-K-3} with
\begin{eqnarray}
\bu_1^\prime &=&
\Delta_1\left(\frac{\bx^\prime}{\phi^\prime}\right)=
\theta_{(1)}\theta_{(12)}\bu_1, \qquad \\
\bu_2^\prime &= -&\Delta_2\left(\frac{\bx^\prime}{\phi^\prime}\right)=
\theta_{(2)}\theta_{(12)}\bu_2,
\end{eqnarray}
and
\begin{equation}
a_i^\prime = \frac{1}{F_{(i)}^\prime \phi_{(i)}^\prime} -
\frac{1}{\phi^\prime}= -\theta_i^2 a_i, \qquad
b_i^\prime= \frac{F^\prime}{\phi_{(i)}^\prime}-\frac{1}{\phi^\prime} = 
-\frac{\theta_{1}\theta_{2}\theta_{12}}{\theta} b_i \quad i=1,2.
\end{equation}
The double points on the line $t_1^\prime=0$ of the second involution
are given by
\begin{equation}
\sqrt{a_1^\prime b_1^\prime}\bu_2^\prime \pm \sqrt{a_2^\prime
b_2^\prime}\bu_1^\prime=
\frac{(\theta_{(1)}\theta_{(2)}\theta_{(12)})^{3/2}}{(\theta)^{1/2}}
(\sqrt{a_1b_1}\bu_2 \pm \sqrt{a_2b_2}\bu_1),
\end{equation}
and due to Theorem \ref{th:inv-dp}, both evolutions are the same.
\end{proof}
\begin{Prop}
Given Koenigs lattice $\bx$ and its transform $\bx^\prime$ then any pair of
intersecting conics of two pencils determines a pencil of quadrics. Such a
pencil defines on the line $L_{\bx\bx^\prime}$ of the congruence the
involution described in Proposition \ref{prop:invol}.
\end{Prop}
\begin{proof}
Let us choose points $\bx$, $\bu_1$, $\bu_2$ and $\bx^\prime$ as the
vertices of the reference frame, then $(t_1,t_2,t_3,t_4)$ are
coordinates of a point
\begin{equation}
\by = t_1\bx + t_2 \bu_1 + t_3 \bu_2 + t_4 \bx^\prime.
\end{equation}
Notice the following relation between these coordinates and the coordinates
$t_i^\prime$ used in the last part of the proof above 
\begin{equation} \label{eq:inv-1-2-transf}
t_1^\prime = t_4, \qquad
t_2^\prime=\frac{t_2}{\theta_{(1)}\theta_{(12)}}, \qquad 
t_3^\prime=-\frac{t_3}{\theta_{(2)}\theta_{(12)}}.
\end{equation}
The transformation formulas \eqref{eq:inv-1-2-transf} imply
that the equation of the second pencil reads
\begin{eqnarray} \label{eq:pencil'}
& \frac{\theta_{(1)}\theta_{(2)}}{\theta\theta_{(12)}}
\left[a_1b_1t_2^2 + a_2b_2 t_3^2 - t_2t_3 \left( 
\frac{\theta_{(1)}}{\theta_{(2)}}a_1b_2 + 
\frac{\theta_{(2)}}{\theta_{(1)}}a_2 b_1 + 
\frac{\mu^\prime\theta}{\theta_{(1)}^2\theta_{(2)}^2\theta_{(12)}}\right)
 \right]+& \nonumber \\
 & - t_2 t_4 \left( \frac{\theta_{(1)}}{\theta_{(12)}}a_1 +
\frac{\theta_{(2)}}{\theta}b_1\right) +
t_3 t_4 \left( \frac{\theta_{(2)}}{\theta_{(12)}}a_2 +
\frac{\theta_{(1)}}{\theta}b_2\right) + t_4^2 =0,
\end{eqnarray}
and, therefore, the relation between parameters of the corresponding
conics in the pencils reads
\begin{equation}  \label{eq:mu-mu'}
\frac{\phi}{\theta_{(2)}}a_1b_2 + 
\frac{\phi}{\theta_{(1)}}a_2 b_1 + 
\frac{\mu^\prime\theta}{\theta_{(1)}^2\theta_{(2)}^2\theta_{(12)}} +\mu = 0.
\end{equation} 
Any two corresponding conics of the pencils \eqref{eq:pencil-K-3} and
\eqref{eq:pencil'} define a pencil of quadrics and, finally,  
we obtain the following two-parameter linear system of quadrics  
\begin{eqnarray} \label{eq:2pls-Q}
&t_1^2 + a_1b_1t_2^2 + a_2b_2 t_3^2 + t_1t_2(a_1+b_1) + t_1t_3(a_2+b_2) +
t_2t_3 (a_1b_2 + a_2 b_1 + \mu) + & \nonumber \\
&+\frac{\theta \theta_{(12)}}{\theta_{(1)}\theta_{(2)}}t_4^2 + 
\nu t_1 t_4 - t_2 t_4 \left( \frac{\theta}{\theta_{(2)}}a_1 +
\frac{\theta_{(12)}}{\theta_{(1)}}b_1\right) +
t_3 t_4 \left( \frac{\theta}{\theta_{(1)}}a_2 +
\frac{\theta_{(12)}}{\theta_{(2)}}b_2\right) = 0.&
\end{eqnarray}
Notice that any pair of two conics of the pencils (i.e., an arbitrary fixed
$\mu$ in equation \eqref{eq:2pls-Q}) determines the same involution on 
the line $t_2=t_3=0$ (the line of the congruence)
\begin{equation}
t_1^2 + \nu t_1 t_4 + \frac{\theta
\theta_{(12)}}{\theta_{(1)}\theta_{(2)}}t_4^2 =0.
\end{equation} 
The fixed points of the involution correspond to 
\begin{equation}
\nu = \pm \sqrt{\frac{\theta
\theta_{(12)}}{\theta_{(1)}\theta_{(2)}}}, 
\end{equation}
and read
\begin{equation} \label{eq:fixed-x-xK}
\pm\sqrt{\theta\theta_{(12)}}\bx + 
\sqrt{\theta_{(1)}\theta_{(2)}}\bx^\prime. 
\end{equation}
One can check using equations \eqref{eq:fixed-x-xK} and 
\eqref{eq:x-xK-focal} that the pairs 
$\bx$ and $\bx^\prime$, $\by_1$ and $\by_{1(1)}$, and
$\by_2$ and $\by_{2(2)}$ are harmonically conjugate with
respect to the fixed points.  
\end{proof}
\begin{Cor} Equations \eqref{eq:pencil-K-F}, \eqref{eq:mu-lambda}, their
primed versions and equation \eqref{eq:mu-mu'} imply that
the distinguished six-point conics of two pencils intersect.
\end{Cor}

\section{The permutability of superpositions of the discrete Koenigs 
transformations}
\label{sec:K-superposition}
We show in this section that superpositions of the discrete 
Koenigs transformations satisfy the permutability property.
We start with the relevant properties of the Moutard equation and then we
present the permutability theorem for the discrete Koenigs transformation
proving this way the integrability of the Koenigs lattice.

\subsection{The discrete Moutard transformation and its permutability
property}
We recall the known material on the Darboux type transformation for 
the Moutard equation \cite{NiSchief,Schief-SDE-Tz} and the permutability
theorem for this transformation \cite{W-cong,NieszporskiA}. The novelty here
is the presentation of the Moutard transformation within the general setting
of the fundamental transformation of quadrilateral lattices.
 
Consider the Moutard lattice, i.e., the quadrilateral lattice
whose homogeneous coordinates $\by:\ZZ^2\to\RR^{N+1}_*$ 
satisfy (up to a gauge) the discrete Moutard equation 
\eqref{eq:Moutard-d}
\begin{equation} 
\by_{(12)} + \by = F (\by_{(1)} + \by_{(2)}) .
\end{equation}
Let $\theta$  be a scalar solution of this equation, linearly
independent of the components of $\by$. 
One can check that the function $\psi=\theta_{(-1)}+\theta_{(-2)}$ satisfies
the equation
\begin{equation} \label{eq:Moutard-d-adj}
\psi_{(12)} + \psi = F_{(-2)} \psi_{(1)} +  F_{(-1)} \psi_{(2)} ,
\end{equation}
adjoint to the Moutard equation. Notice that, like in the Koenigs
lattice case, to construct the fundamental
transformation of the Moutard lattice we need only half of the data, 
but this time the solution of the
Moutard equation gives a solution of its adjoint.
The solution of the system  \eqref{eq:kl} (with the change of notation
$\phi\to\theta$ and $\theta\to\psi$) reads then 
\begin{equation}
k=\theta\theta_{(-1)}, \qquad \ell=-\theta\theta_{(-2)},
\end{equation}
and the lattice $\hat\by$, the solution of the linear system
\begin{eqnarray}
\Delta_{1}(\hat\by) & =& \theta\theta_{(1)} 
\Delta_{1}\left(\frac{\by}{\theta}\right),\\ 
\Delta_{2}(\hat\by) & =& - \theta\theta_{(2)} 
\Delta_{2}\left(\frac{\by}{\theta}\right),
\end{eqnarray}
satisfies the Laplace equation
\begin{equation}
\hat\by_{(12)}=F\frac{\theta_{(2)}}{\theta}\hat\by_{(1)} +
F\frac{\theta_{(1)}}{\theta}\hat\by_{(2)} + 
\left(1- F\frac{\theta_{(2)}}{\theta}-F\frac{\theta_{(1)}}{\theta}\right)
\hat\by.
\end{equation}
Its gauge transform $\by^\prime$ defined by
\begin{equation}
\hat\by=\by^\prime \theta
\end{equation}
satisfies the Moutard equation with the potential $F^\prime$ given by
equation \eqref{eq:F'}.

Finally, we obtain the known \cite{Schief-SDE-Tz} formulas
\begin{eqnarray}
\label{eq:dmt}
\Delta_{1}(\theta\by^\prime) & =&  \theta\theta_{(1)} 
\Delta_{1}\left(\frac{\by}{\theta}\right),\\ 
\Delta_{2}(\theta\by^\prime) & =& - \theta\theta_{(2)} 
\Delta_{2}\left(\frac{\by}{\theta}\right),
\end{eqnarray}
which allow to find the new Moutard lattice $\by'$ given the old Moutard
lattice $\by$ and the scalar
solution $\theta$ of the Moutard equation of $\by$. Notice that
$\theta^\prime=1/\theta$ satisfies the Moutard equation of $\by^\prime$.

Let $\theta^1$ and $\theta^2$ be two solutions of the Moutard equation
\eqref{eq:Moutard-d}. Denote by $\by^{(1)}$ and $\theta^{2(1)}$
the transforms of $\by$ and $\theta^2$ via
$\theta^1$ and denote by $\by^{(2)}$ and $\theta^{1(2)}$
the transforms of $\by$ and $\theta^1$  via $\theta^2$. Then $\by^{(1)}$, 
$\theta^{2(1)}$ and $\theta^{1(1)}=1/\theta^1$ satisfy the Moutard equation
with the potential
\begin{equation} \label{eq:F1}
F^{(1)}=F\frac{\theta^1_{(1)}\theta^1_{(2)}}{\theta^1\theta^1_{(12)}},
\end{equation}
and 
$\by^{(2)}$, $\theta^{1(2)}$ and $\theta^{2(2)}=1/\theta^2$
satisfy the Moutard equation
with the potential 
\begin{equation} \label{eq:F2}
F^{(2)}=F\frac{\theta^2_{(1)}\theta^2_{(2)}}{\theta^2\theta^2_{(12)}}.
\end{equation}
Notice \cite{W-cong,NieszporskiA}  that the transformation formulas 
\eqref{eq:dmt} give
\begin{eqnarray}
\label{eq:m12}
\Delta_1(\theta^1\theta^{2(1)})&=&-\Delta_1(\theta^2\theta^{1(2)}),\\
\Delta_2(\theta^1\theta^{2(1)})&=&-\Delta_2(\theta^2\theta^{1(2)}),
\end{eqnarray}
which implies that fixing one of the two integration constants we have
\begin{equation} \label{eq:Xi}
\theta^1\theta^{2(1)}=-\theta^2 \theta^{1(2)} = \Xi.
\end{equation}
Then the lattices
$\by^{(12)}$ of the one parameter family 
(due to the an additive constant in $\Xi$) given by 
\begin{equation} \label{eq:yM12}
\by^{(12)} = \by + \frac{\theta^1\theta^2}{\Xi}(\by^{(1)}-\by^{(2)}),
\end{equation}
are
simultaneously transforms of $\by^{(1)}$ via $\theta^{2(1)}$ and 
transforms of $\by^{(2)}$ via $\theta^{1(2)}$.
\begin{Rem}
To obtain symmetric more form of the superposition formula \eqref{eq:yM12} 
one can use the allowed gauge freedom in the transformation formulas
\cite{W-cong,NieszporskiA}.
\end{Rem}
 
\subsection{Superposition of the discrete Koenigs transformations}
Let us use $\theta^1$ and $\theta^2$ to find two transforms of the Koenigs
lattice $\bx$ satisfying equation \eqref{eq:Koenigs-d}. According to
notation of Proposition \ref{th:transf-dK} denote by
$\phi^1=\theta^1_{(1)} + \theta^1_{(2)}$ and
$\phi^2=\theta^2_{(1)} + \theta^2_{(2)}$ the corresponding solutions of
the Koenigs lattice equation.  
Denote by $\bx^{(1)}$ and $\phi^{2(1)}$ 
the transforms of $\bx$ and $\phi^2$ with respect to $\theta^1$, i.e.,
\begin{eqnarray} \label{eq:dK-x1-1}
\Delta_1\left( \frac{1}{\phi^{1(1)}} 
\left( \begin{array}{c} \bx^{(1)} \\ \phi^{2(1)} \end{array} \right)
\right) & = &\; 
(\theta^1 \theta^1_{(2)})_{(1)}
\Delta_1\left( \frac{1}{\phi^1} 
\left( \begin{array}{c} \bx \\ \phi^{2} \end{array} \right)
\right), \\
\Delta_2\left( \frac{1}{\phi^{1(1)}} 
\left( \begin{array}{c} \bx^{(1)} \\ \phi^{2(1)} \end{array} \right)
\right) & = &-
(\theta^1 \theta^1_{(1)})_{(2)}
\Delta_2\left( \frac{1}{\phi^1} 
\left( \begin{array}{c} \bx \\ \phi^{2} \end{array} \right)
\right),
\end{eqnarray}
where
\begin{equation}
\phi^{1(1)} = \frac{1}{\theta^1_{(1)}} + \frac{1}{\theta^1_{(2)}}.
\end{equation}
According to Proposition \ref{th:transf-dK} the functions
$\bx^{(1)}$, $\phi^{2(1)}$  and 
$\phi^{1(1)}$ satisfy the Koenigs lattice equation with
the transformed potential $F^{(1)}$ given by \eqref{eq:F1}.
Similarly, by $\bx^{(2)}$ and $\phi^{1(2)}$ denote
the transforms of $\bx$ and $\phi^1$
with respect to $\theta^2$, i.e.,
\begin{eqnarray}
\Delta_1\left( \frac{1}{\phi^{2(2)}} 
\left( \begin{array}{c} \bx^{(2)} \\ \phi^{1(2)} \end{array} \right)
\right) & = &\; 
(\theta^2 \theta^2_{(2)})_{(1)}
\Delta_1\left( \frac{1}{\phi^2} 
\left( \begin{array}{c} \bx \\ \phi^{1} \end{array} \right)
\right), \\
\Delta_2\left( \frac{1}{\phi^{2(2)}} 
\left( \begin{array}{c} \bx^{(2)} \\ \phi^{1(2)} \end{array} \right)
\right) & = &-
(\theta^2 \theta^2_{(1)})_{(2)}
\Delta_2\left( \frac{1}{\phi^2} 
\left( \begin{array}{c} \bx \\ \phi^{1} \end{array} \right)
\right),
\end{eqnarray}
where
\begin{equation} \label{eq:phi2-2}
\phi^{2(2)} = \frac{1}{\theta^2_{(1)}} + \frac{1}{\theta^2_{(2)}},
\end{equation}
and $\,\bx^{(2)}$, $\phi^{1(2)}$  and 
$\phi^{2(2)}$ satisfy the Koenigs lattice equation with
the transformed potential given by \eqref{eq:F2}.

\begin{Prop}
The lattices $\bx^{(12)}$ of the one parameter family 
(because of the free parameter in the definition of
$\Xi$) given by
\begin{equation} \label{eq:x-sup-K}
\bx^{(12)} = 
-\frac{\Xi_{(1)}\Xi_{(2)}\phi^{1(21)}\phi^{2(12)}}{\phi^1\phi^2}\bx
+\frac{\phi^{1(21)}}{\phi^{1(1)}}\bx^{(1)} +
\frac{\phi^{2(12)}}{\phi^{2(2)}}\bx^{(2)},
\end{equation}
where
\begin{equation}
\phi^{1(21)}=\frac{1}{\theta^{1(2)}_{(1)}}+\frac{1}{\theta^{1(2)}_{(2)}}, \qquad
\phi^{2(12)}=\frac{1}{\theta^{2(1)}_{(1)}}+\frac{1}{\theta^{2(1)}_{(2)}},
\end{equation}
are simultaneously the Koenigs transforms of $\bx^{(1)}$ via $\theta^{2(1)}$ 
and the Koenigs transforms of $\bx^{(2)}$ via $\theta^{1(2)}$.
\end{Prop}
\begin{proof}
We have to check that $\bx^{(12)}=\bx^{(21)}$ defined in 
\eqref{eq:x-sup-K} satisfies equations
\begin{eqnarray}
\Delta_1\left( \frac{\bx^{(12)}}{\phi^{2(12)}} \right) & = &\;
 (\theta^{2(1)} \theta^{2(1)}_{(2)})_{(1)}
 \Delta_1\left( \frac{\bx^{(1)}}{\phi^{2(1)}} \right), 
 \label{eq:K-sup-12-1}\\
\Delta_2\left( \frac{\bx^{(12)}}{\phi^{2(12)}} \right) & = &
- (\theta^{2(1)} \theta^{2(1)}_{(1)})_{(2)}
\Delta_2\left( \frac{\bx^{(1)}}{\phi^{2(1)}} \right),
\label{eq:K-sup-12-2}
\end{eqnarray}
which define the 
transform $\bx^{(12)}=(\bx^{(1)})^{(2)}$ of $\bx^{(1)}$ via
$\theta^{2(1)}$, and satisfies equations
\begin{eqnarray}
\Delta_1\left( \frac{\bx^{(21)}}{\phi^{1(21)}} \right) & = &\;
 (\theta^{1(2)} \theta^{1(2)}_{(2)})_{(1)}
 \Delta_1\left( \frac{\bx^{(2)}}{\phi^{1(2)}} \right), 
 \label{eq:K-sup-21-1}\\
\Delta_2\left( \frac{\bx^{(21)}}{\phi^{1(21)}} \right) & = &
- (\theta^{1(2)} \theta^{2(2)}_{(1)})_{(2)}
\Delta_2\left( \frac{\bx^{(2)}}{\phi^{1(2)}} \right),
\label{eq:K-sup-21-2}
\end{eqnarray}
which define the transform $\bx^{(21)}=(\bx^{(2)})^{(1)}$ of $\bx^{(2)}$ via
$\theta^{1(2)}$. This can be done by direct verification using equations
\eqref{eq:m12}--\eqref{eq:Xi} and \eqref{eq:dK-x1-1}--\eqref{eq:phi2-2}.
\end{proof}
\begin{Rem}
To obtain the superposition formula \eqref{eq:x-sup-K} we assume that
$\bx^{(21)}=\bx^{(12)}$ and
formulas \eqref{eq:K-sup-12-1} -- \eqref{eq:K-sup-21-2} hold.
\end{Rem}

\bibliographystyle{amsplain}

\begin{thebibliography}{10}

\bibitem{Bianchi}
L.~Bianchi, \emph{Lezioni di geometria differenziale}, Zanichelli, Bologna,
  1924.

\bibitem{BobenkoSeiler}
A.~Bobenko and R.~Seiler (eds.), \emph{Discrete integrable geometry and
  physics}, Clarendon Press, Oxford, 1999.

\bibitem{SIDEII}
P.~Clarkson and F.~Nijhoff (eds.), \emph{{S}ymmetries and {I}ntegrability of
  {D}ifference {E}quations}, University Press, Cambridge, 1999.

\bibitem{DarbouxIV}
G.~Darboux, \emph{Le\c{c}ons sur la th\'{e}orie g\'{e}n\'{e}rale des surfaces.
  {I--IV}}, Gauthier -- Villars, Paris, 1887--1896.

\bibitem{DKJM}
E.~Date, M.~Kashiwara, M.~Jimbo, and T.~Miwa, \emph{Transformation groups for
  soliton equations}, Nonlinear integrable systems --- classical theory and
  quantum theory, Proc. of RIMS Symposium (M.~Jimbo and T.~Miwa, eds.), World
  Scientific, Singapore, 1983, pp.~39--119.

\bibitem{DCN}
A.~Doliwa, \emph{Geometric discretisation of the {Toda} system}, Phys. Lett. A
  \textbf{234} (1997), 187--192.

\bibitem{q-red}
\bysame, \emph{Quadratic reductions of quadrilateral lattices}, J. Geom. Phys.
  \textbf{30} (1999), 169--186.

\bibitem{W-cong}
\bysame, \emph{Discrete asymptotic nets and {W}--congruences in {P}l\"ucker
  line geometry}, J. Geom. Phys. \textbf{39} (2001), 9--29.

\bibitem{DMS}
A.~Doliwa, S.~V. Manakov, and P.~M. Santini, \emph{$\bar\partial$-reductions of
  the multidimensional quadrilateral lattice: the multidimensional circular
  lattice}, Comm. Math. Phys. \textbf{196} (1998), 1--18.

\bibitem{DMMMS}
A.~Doliwa, M.~Ma{\~n}as, L.~Mart{\'\i}nez Alonso, E.~Medina, and P.~M. Santini,
  \emph{Multicomponent {KP} hierarchy and classical transformations of
  conjugate nets}, J. Phys. A \textbf{32} (1999), 1197--1216.

\bibitem{DNS-I}
A.~Doliwa, M.~Nieszporski, and P.~M. Santini, \emph{Asymptotic lattices and
  their integrable reductions: {I}. the {B}ianchi--{E}rnst and the
  {F}ubini--{R}agazzi lattices}, J. Phys A: Math. Gen. \textbf{34} (2001),
  10423--10439.

\bibitem{MQL}
A.~Doliwa and P.~M. Santini, \emph{Multidimensional quadrilateral lattices are
  integrable}, Phys. Lett. A \textbf{233} (1997), 365--372.

\bibitem{DS-sym}
\bysame, \emph{The symmetric, {D}-invariant and {E}gorov reductions of the
  quadrilateral lattice}, J. Geom. Phys. \textbf{36} (2000), 60--102.

\bibitem{TQL}
A.~Doliwa, P.~M. Santini, and M.~Ma{\~n}as, \emph{Transformations of
  quadrilateral lattices}, J. Math. Phys. \textbf{41} (2000), 944--990.

\bibitem{Eisenhart-K}
L.~P. Eisenhart, Annals of Mathematics \textbf{18} (1916), 11.

\bibitem{Eisenhart-TS}
\bysame, \emph{Transformations of surfaces}, Princeton University Press,
  Princeton, 1923.

\bibitem{Finikov}
S.~P. Finikov, \emph{Theorie der {K}ongruenzen}, Akademie-Verlag, Berlin, 1959.

\bibitem{JM}
M.~Jimbo and T.~Miwa, \emph{Algebraic analysis of solvable lattice models},
  Regional Conference Series in Mathematics, vol.~85, AMS, Providence, 1995.

\bibitem{KvL}
V.~G. Kac and J.~van~de Leur, \emph{The n-component {KP} hierarchy and
  representation theory}, Important developments in soliton theory (A.~S. Fokas
  and V.~E. Zakharov, eds.), Springer, Berlin, 1993, pp.~302--343.

\bibitem{Koenigs}
G.~Koenigs, Comptes Rendus, Acad\'emie des Sciences de l'Institut de France
  \textbf{113} (1891), 1022.

\bibitem{KoSchief2}
B.~G. Konopelchenko and W.~K. Schief, \emph{Three-dimensional integrable
  lattices in {Euclidean} spaces: Conjugacy and orthogonality}, Proc. Roy. Soc.
  London A \textbf{454} (1998), 3075--3104.

\bibitem{KBI}
V.~E. Korepin, N.~M. Bogoliubov, and A.~G. Izergin, \emph{Quantum inverse
  scattering method and correlation functions}, Cambridge Univ. Press,
  Cambridge, 1993.

\bibitem{Lane}
E.~P. Lane, \emph{Projective differential geometry of curves and surfaces},
  Univ. Chicago Press, Chicago, 1932.

\bibitem{SIDEIII}
D.~Levi and O.~Ragnisco (eds.), \emph{{SIDE} {III} -- {S}ymmetries and
  {I}ntegrability of {D}ifference {E}quations}, CMR Proceedings and Lecture
  Notes, vol.~25, AMS, Providence, 2000.

\bibitem{SIDEI}
D.~Levi, L.~Vinet, and P.~Winternitz (eds.), \emph{Symmetries and
  {I}ntegrability of {D}ifference {E}quations}, CMR Proceedings and Lecture
  Notes, vol.~9, AMS, Providence, 1996.

\bibitem{MDS}
M.~Ma{\~n}as, A.~Doliwa, and P.~M. Santini, \emph{Darboux transformations for
  multidimensional quadrilateral lattices. {I}}, Phys. Lett. A \textbf{232}
  (1997), 99--105.

\bibitem{Nieszporski-priv-A01}
M.~Nieszporski, private communication, August 2001.

\bibitem{NieszporskiA}
\bysame, \emph{On discretization of asymptotic nets}, J. Geom. Phys.
  \textbf{40} (2001), 259--276.

\bibitem{NDS}
M.~Nieszporski, A.~Doliwa, and P.~M. Santini, \emph{The integrable
  discretization of the {B}ianchi--{E}rnst system}, nlin.{SI}/0104065.

\bibitem{NiSchief}
J.~J.~C. Nimmo and W.~K. Schief, \emph{Superposition principles associated with
  the {Moutard} transformation. {An} integrable discretisation of a
  (2+1)-dimensional sine-{Gordon} system}, Proc. R. Soc. London A \textbf{453}
  (1997), 255--279.

\bibitem{Samuel}
P.~Samuel, \emph{Projective geometry}, Springer, New York--Berlin--Heidelberg,
  1988.

\bibitem{Sauer}
R.~Sauer, \emph{Differenzengeometrie}, Springer, Berlin, 1970.

\bibitem{Schief-SDE-Tz}
W.~K. Schief, \emph{Self-dual {E}instein spaces and a discrete {T}zitzeica
  equation. a permutability theorem link}, Symmetries and Integrability of
  Difference Equations (P.~A. Clarkson and F.~W. Nijhoff, eds.), University
  Press, Cambridge, 1999, pp.~137--148.

\bibitem{Tzitzeica-K}
G.~Tzitz\'eica, Comptes Rendus, Acad\'emie des Sciences de l'Institut de France
  \textbf{147} (1908), 1036.

\end{thebibliography}

\providecommand{\bysame}{\leavevmode\hbox to3em{\hrulefill}\thinspace}

\end{document}